\documentclass[preprint2]{aastex}

\newcommand{\simgt}{\,\rlap{\lower 3.5 pt \hbox{$\mathchar \sim$}} \raise
1pt \hbox {$>$}\,}
\newcommand{\simlt}{\,\rlap{\lower 3.5 pt \hbox{$\mathchar \sim$}} \raise
1pt \hbox {$<$}\,}

\def \kms {\ifmmode  \,\rm km\,s^{-1} \else $\,\rm km\,s^{-1}  $ \fi }
\def \kpc {\ifmmode  {\rm~kpc}  \else ${\rm~kpc}$\fi}
\def \pc {\ifmmode  {\rm~pc}  \else ${\rm~pc}$ \fi  }
\def \Gyr {\ifmmode  {\rm~Gyr}  \else ${\rm~Gyr}$\fi}
\def \Msun {\ifmmode M_{\odot} \else $M_{\odot}$ \fi}
\def \Lsun {\ifmmode L_{\odot} \else $L_{\odot}$ \fi}
\def \Rsun {\ifmmode R_{\odot} \else $R_{\odot}$ \fi}
\def \Msunpyr {\ifmmode M_{\odot}{\rm~yr}^{-1} \else $M_{\odot}{\rm~yr}^{-1}$ \fi}
 \def \hMsun {\ifmmode h^{-1}\,\rm M_{\odot} \else $h^{-1}\,\rm M_{\odot}$ \fi}

\shorttitle{Temperature and velocity anistropy}
\shortauthors{Hansen et al.}

\begin{document}


\title{The temperature of hot gas in galaxies and clusters:\\
baryons dancing to the tune of dark matter}


\author{Steen H. Hansen$^1$, Andrea V. Macci\'o$^2$, Emilio Romano-Diaz$^3$,
   Yehuda Hoffman$^4$, Marcus Br{\"u}ggen$^5$, Evan Scannapieco$^6$,
and Greg S. Stinson$^7$}

\affil{$^1$ Dark Cosmology Centre, Niels Bohr Institute, University of Copenhagen,\\
Juliane Maries Vej 30, 2100 Copenhagen, Denmark}
\affil{$^2$ Max-Planck-Institut f\"ur Astronomie, K\"onigstuhl 17, 69117 Heidelberg, Germany }
\affil{$^3$Department of Physics \& Astronomy, University of Kentucky, Lexington, KY 40506, USA}
\affil{$^4$Racah Institute of Physics, Hebrew University, Jerusalem 91904, Israel}
\affil{$^5$Jacobs University Bremen, P.O. Box 750\,561, 28725 Bremen, Germany}
\affil{$^6$School of Earth and Space Exploration,  Arizona State University, P.O.  Box 871404, Tempe, AZ, 85287-1404, USA}
\affil{$^7$Jeremiah Horrocks Institute, University of Central Lancashire, Preston PR1 2HE}



\begin{abstract}

  The temperature profile of hot gas in galaxies and galaxy clusters
  is largely determined by the depth of the total gravitational
  potential and thereby by the dark matter (DM) distribution. We
  use high-resolution hydrodynamical simulations of galaxy formation
  to derive a surprisingly simple relation between the gas
  tempe\-ra\-ture and DM properties. We show that this relation holds
  not just for galaxy clusters but also for equilibrated and relaxed
  galaxies at radii beyond the central stellar-dominated region of
  typically a few kpc.  It is then clarified how a measurement of the
  temperature and density of the hot gas component can lead to an indirect
  measurement of the DM velocity anisotropy in galaxies.  We also
  study the temperature relation for galaxy clusters in the presence
  of self-regulated, recurrent active galactic nuclei (AGN), and
  demonstrate that this temperature relation even holds outside the
  inner region of $\approx 30$ kpc in clusters with an active AGN.

\end{abstract}




\section{Introduction}

Dark matter (DM) dominated cosmological structures are a direct
outcome of hierarchical structure formation with a subdominant baryon
fraction.  The baryons can either cool fast to form stars or end up as
a hot virialized gas
\citep{1977MNRAS.179..541R,1977ApJ...211..638S,2003MNRAS.345..349B,2005MNRAS.363....2K}. This
hot gas is customarily observed in galaxies and galaxy clusters
\citep{1986RvMP...58....1S}, and can be heated through 
adiabatic compression, shock heating or
non-gravitational processes and cooled through radiative processes.


The ability of the hot gas to shock heat to the
virial temperature opens the possibility that one can combine the
equations governing the dynamics of the gas and the DM, namely the
equation of hydrostatic equilibrium and the Jeans
equation.
The
simultaneous solution to these equations can in principle allow us to
determine the DM velocity dispersion anisotropy, which holds
information about the fundamental difference in the way DM and baryons
equilibrate in cosmological structures. In fact, such a method has
already been applied to galaxy clusters, where numerical simulations
of cluster formation and evolution have been used to confirm that the
gas equilibrium temperature is determined through the averaged
velocity dispersion of the DM~\citep{hansenpiff,host2009}.  The
resolution of these simulations, however, did not allow us to probe radii
smaller than a few hundred kpc. The temperature of the cooling gas
remains to be determined at smaller radii and in the presence of an
active AGN.  Furthermore, it remains unknown if the gas temperature in
galaxies will obey a similar simple relation.

In this paper we conduct a dedicated comparison between the gas and DM
in a range of simulated cosmological objects.  The
structure of the paper is as follows: first we discuss the relationship
between the gas temperature and the DM velocity dispersion.  In
section \ref{sec:num} we present the results of numerical simulations
of galaxy formation, and confirm that the gas temperature in galaxies
exhibits a bimodal distribution --- the hot gas resides at the DM
temperature and the cold gas cools below $10^4$~K and forms stars.  In section
\ref{sec:agn} we present the results of numerical simulations of AGN
outflows, which demonstrate that even when the gas is heated
episodically by a self-regulated AGN, it still moves towards the
DM temperature, which it succeeds in reaching already beyond $\approx
30$~kpc.
Finally, in section~\ref{sec:discussion} we explain how our findings open up
for the possibility of measuring the DM velocity anisotropy in
galaxies, and in section \ref{sec:conclusion} we briefly offer our
conclusions.

\section{The temperature relation}

It has been known for years that baryons in a DM-dominated potential 
will heat up to a temperature that is determined by the properties of 
the DM mass profile \citep{1977MNRAS.179..541R, 1978A&A....70..677C, 
1981A&A...100..194C, 1986RvMP...58....1S,  2009ApJ...698..580C}~\footnote{The corresponding stellar velocity dispersion will also
be of the same magnitude as the DM velocity dispersion.
However, there are no
theoretical reason why the anisotropies of the galaxies should
be the same as that of the DM~\citep{2003MNRAS.343..401L,radek}.}. This is easily seen
when considering the similarity between the equation of hydrostatic 
equilibrium \citep{1986RvMP...58....1S} and the Jeans equation 
\citep{binneytremaine}. The former depends on the
gas temperature, $T$, and through its equation of state 
on the density, $\rho_{\rm gas}$, while the latter 
depends on the DM density, $\rho_{\rm DM}$, its radial and tangential 
velocity dispersions, $\sigma_{\rm r,t}^2$, and the velocity anisotropy
$\beta = 1-\sigma_t^2 /\sigma^2_r$,
\begin{eqnarray}
\frac{GM_{\rm tot}}{r} &=& - \frac{T k_B}{\mu m_p} \,
\left[
\frac{d {\rm log} \rho_{\rm gas}}{d {\rm log}r} +
\frac{d {\rm log} T}{d {\rm log}r}
\right] \, 
\label{eq:hydro} \\
\frac{GM_{\rm tot}}{r} &=& - \sigma^2_r \,
\left[
\frac{d {\rm log} \rho_{\rm DM}}{d {\rm log}r} +
\frac{d {\rm log} \sigma^2_r}{d {\rm log}r} + 2 \beta
\right] \, ,
\label{eq:jeans}
\end{eqnarray}
where
$\mu m_p$ is the averaged mass of the gas particles.
For a given total mass profile, the hydrostatic equilibrium
equation, eq.~(\ref{eq:hydro}), has a wide range of possible
solutions: basically any radial gas temperature profile is 
allowed (in principle), since a gas density profile can always
be constructed to fulfill the equation. 
Furthermore, there is a priori no theoretical connection between 
the gas temperature and the dark matter dispersions, except that the
particles are sitting in the same gravitational potential.
One can, however, use the dark matter dispersions to define
a {\em ``DM temperature''} by
\begin{equation}
\frac{T_{\rm DM} k_B}{\mu m_p} \equiv \frac{1}{3} \left( 
\sigma^2_{\rm r} + \sigma^2_{\rm \theta} + \sigma^2_{\rm \phi} 
\right) \, ,
\label{eq:sigtot}
\end{equation}
where $\sigma_{\rm \theta}$ and $\sigma_{\rm \phi}$ are the DM
tangential dispersion velocities. 
We emphasize that there is no fundamental reason why the
temperature of the gas, $T$, should be identical to
the DM temperature (in praxis the DM only sets an upper limit to the
gas temperature). We will, however, consider the
ratio between the gas temperature, and the temperature 
derived from the DM dispersions, namely
\begin{equation}
\kappa (r) \equiv \frac{T k_B}{\mu m_p}
\frac{3}{\sigma^2_{\rm r} + \sigma^2_{\rm \theta} + \sigma^2_{\rm \phi}} \, ,
\label{eq:kappadef}
\end{equation}
and we will discuss a possible universality of this
ratio. There is, to our knowledge, no simple dynamical argument
why this ratio should be close to unity, or why it should have
a universal radial behaviour. For instance, in the adiabatic
simulations presented in \cite{2007MNRAS.376.1327F} it was found
that $\kappa$ decreases with radius in the inner region. We will
here address this question using galaxy formation including 
cooling, heating and feedback.

Naturally,
gravity does not care about the masses of the individual
(collisionless) particles, which is why their velocity dispersions are
equal, and not their temperatures.  The standard definition of the
temperature is written in terms of the average kinetic energy of
particles and describes a system in thermodynamic equilibrium. This
thermodynamic limit is not achieved for most particle systems with
long-range forces such as gravitational structures, for which this
steady state is normally not described by simple distribution
functions, e.g., Maxwell-Boltzmann statistics. Thus, the DM is not in
thermal equilibrium, and technically it does not have a ``temperature'', 
however,  the use of this word should not
cause any confusion.



An important point
is that the DM temperature, as defined in
eq.~(\ref{eq:sigtot}), is not only a function of the mass
profile or potential
of the DM, but also depends on the
dynamical state of the DM through its velocity anisotropy,
$\beta(r)$. The velocity anisotropy  is much harder to observationally
determine than the mass profile.
Hence for a nonvanishing velocity dispersion anisotropy in the DM, the
DM temperature is not identical to the velocity
dispersion of the DM along any random direction, such as the radial
direction, but only to their spatial average.  


We will now proceed to analyse  
the ratio between the local gas and DM temperatures
defined in eq.(\ref{eq:kappadef}),
and use numerical
simulations of galaxy clusters or galaxies to test if
$\kappa$ is of order unity.  Considering the results of two
different numerical codes \citep{kay,2006NewA...12...71V} simulating
both DM and baryonic physics, \cite{host2009} demonstrated that the
$\kappa =1$ within $20 \%$ for relaxed galaxy clusters. 
The gas in galaxies is
much denser, and the cooling times are, therefore, much shorter, and
one might have expected that $\kappa$ might have a much more complicated 
profile.
We demonstrate below that indeed the hot gas in galaxies does have
virtually universal temperature ratios, with $\kappa \approx 1$.

\section{Numerical simulations}
\label{sec:num}

The hydrodynamical simulations were performed with {\sc gasoline}
(Wadsley et al. 2004; see Governato et al. (2007) or Stinson et al. (2010)
for a more detailed
description) --- multi-stepping, parallel Tree\-SPH $N$-body code. We
include radiative and Compton cooling for a primordial mixture of
hydrogen and helium. The star formation algorithm is based on a Jeans
instability criteria (Katz 1992), where gas particles in dense,
unstable regions and in convergent flows spawn star particles at a
rate proportional to the local dynamical time (see Governato et
al. 2004). The star formation efficiency was set to $0.05$, but in the
adopted scheme its precise value has only a minor effect on the star
formation rate (Katz 1992). The code also includes supernova feedback
(Stinton et al. 2006), and a UV background (Haardt \& Madau 1996;
Heller 1999).  Additional simulations using the FTM 4.5 code (Heller
\& Shlosman 1994) are described in Section~3.1.

\begin{figure}[thb]
	\centering
	\includegraphics[angle=0,width=0.45\textwidth]{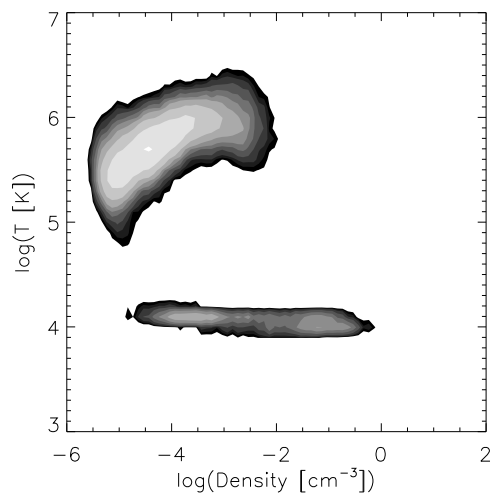}
	\caption{Contour plot of the gas density and temperature of
          the individual gas particle at $z=0$ from the numerical
          simulation using {\sc gasoline} of the galaxy G2. The gas
          particles either cool down to the floor of the atomic
          cooling $10^4$ K, or stay at the DM temperature around
          $10^6$~K.}
\label{fig:phase}
\end{figure}

For the simulations with {\sc gasoline}, we selected three candidate
haloes with masses similar to the mass of the Milky Way ($M \approx
10^{12} \Msun$) from an existing low-resolution DM simulation (300$^3$
particles within 90 Mpc) and re-simulated them at higher resolution.
These high-resolution runs are 8$^3$ times more resolved in mass than
the initial ones and included a gaseous component within the entire
high-resolution region.  In these runs the masses of the DM and gas particles are
respectively $m_{d} = 1.17 \times 10^6 \hMsun$ and $m_g = 2.3 \times
10^5 \hMsun$.  The DM has a spline gravitational softening length of
500 $h^{-1}$ pc and there are about $10^6$ particles for each
component (dark and gas) in the high-resolution region. More
properties of the galaxies are listed in Table \ref{table:haloes}.
These hydrodynamical simulations are described in details in
Schewtschenko \& Macci\`o (2010).


\begin{table}[thb]
 \centering
 \begin{minipage}{140mm}
  \caption{Galaxies parameters}
  \begin{tabular}{lcccc}
\hline  Galaxy &  Mass  &  $R_{\rm vir}$ & $V_{\rm circ}$  \\
       &($10^{12}\hMsun)$&   (kpc~$h^{-1}$)          & (km~s$^{-1}$)   \\
\hline 
G0   & 0.74  & 188 & 178 \\
G1   & 0.89  & 199 & 188  \\
G2   & 0.93  & 202 & 203  \\
\hline 
\label{table:haloes}
\end{tabular}
\end{minipage}
\end{table}

\begin{figure}[thb]
	\centering
	\includegraphics[angle=0,width=0.45\textwidth]{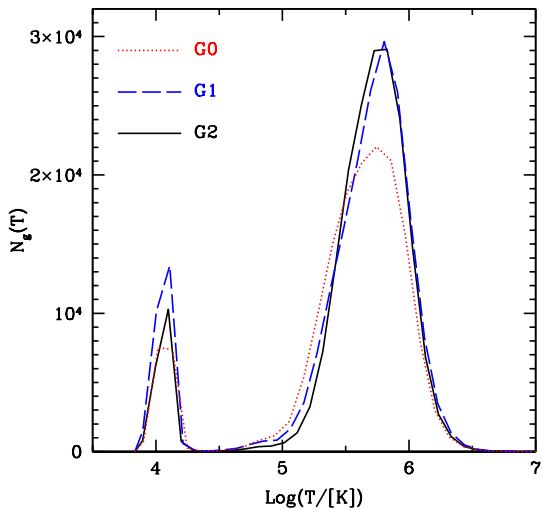}
	\caption{Histogram of the individual gas particle temperatures
          at $z=0$ from the numerical simulation using {\sc
            gasoline}. The three curves represent each of the three
          galaxies simulated at high resolution.  The gas particles
          either cool down to the floor of the atomic cooling $10^4$
          K, or stay at the DM temperature around $10^6$~K.}
\label{fig:tempbump}
\end{figure}

In Fig.~\ref{fig:phase} we show contours over the temperature and
local density of all the gas particles in one of the 3 galaxies (the
others look essentially identical).  In Fig.~\ref{fig:tempbump} we
show histograms over the individual gas particle temperatures in the
three selected galaxies, which is a projection of
figure~\ref{fig:phase}. The gas temperature exhibits a bimodal
distribution: it either cools down to the floor of atomic cooling
about $10^{4}$K in dense clumps (central disk/bulge or satellite
cores) or stays at the DM temperature in the larger and less dense
regions. This is largely caused by the significantly shorter cooling
time around $10^5$K (see e.g. \cite{2005MNRAS.363....2K}). When we
discuss the gas temperature from the simulations using {\sc gasoline}
we refer to the second (large) bump. Practically we calculate a
mass-averaged temperature by including only particles warmer than
$10^{4.5}$K. Making a mass-average over all particles (including the
cold ones) makes a negligible difference at most times. There is no
metal cooling in this simulation, which is the reason for the floor at
$10^{4}$K. The galaxies are typical field galaxies, and have been
selected to have a quiet merger history (except G1, which had a major
merger at $z=0.8$).

\begin{figure}[thb]
	\centering
	\includegraphics[angle=0,width=0.45\textwidth]{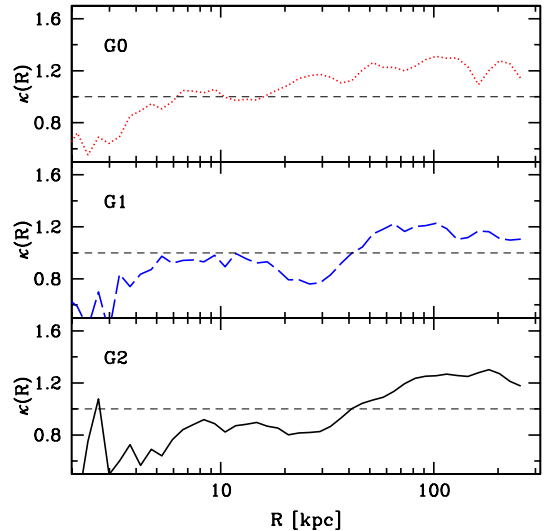}
	\caption{The temperature ratios as a function of radius for
          the three simulated galaxies. The frames show that outside
          the central region of $\approx 6$~kpc where the stellar disk
          dominates the potential, the ratio, $\kappa (r) =
          \sigma^2_{\rm baryons} / \langle \sigma^2_{\rm DM} \rangle$
          approaches unity. }
\label{fig:kappaAM}
\end{figure}

We can now consider radial bins for each galaxy at $z=0$, and in each bin we
calculate the mass-averaged gas temperature and the mass-averaged
DM velocity dispersion. Their ratios as in eq.(\ref{eq:kappadef})  is 
shown in Fig.~\ref{fig:kappaAM}.
From this figure we see that outside the range of the
disk, i.e. beyond $\approx 6$~kpc, these ratios remain close to
unity. Beyond 20~kpc, where the gas density is lower and hence the
cooling time longer, the gas temperature is slightly above the DM
dispersion velocities by 10\%--20\%. The velocity anisotropy is nearly
constant, $\beta (r) \approx 0.2$ outside 2~kpc, slowly increasing to
about $0.3$ near 100~kpc.  We thus confirm that indeed there
remains a substantial component of the hot gas at the DM
temperatures even in dense environments of galaxies.

\begin{figure}[htb]
	\centering
	\includegraphics[angle=0,width=0.45\textwidth]{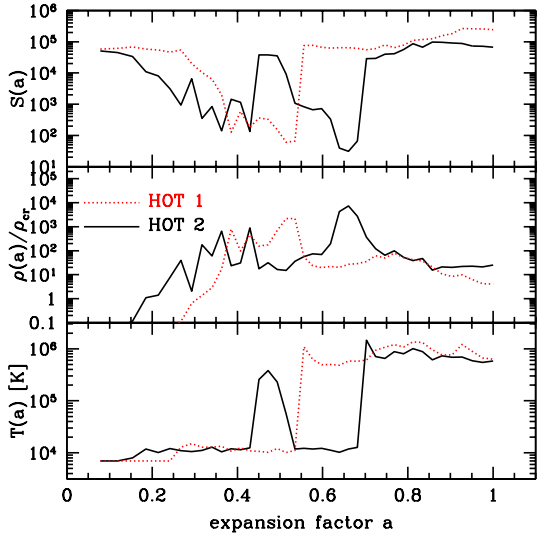}
	\caption{History of 2 hot gas particles. Shown is entropy, local
          density, and temperature as function of expansion factor.}
\label{fig:hothot}
\end{figure}

\subsection{History of hot and cold particles}

In order to better understand the origin of the above bimodal
distribution of gas temperatures, we tracked the evolution of
various properties of individual particles.  Each gas particle has its
own unique thermal history, however, it appears that we as a first
approximation can split the particles into four groups.
\begin{itemize}
\item{\bf 1)} Hot particles, which are shock heated to roughly $10^6$
  K, and then keep that temperature.
\item{\bf 2)} Warm particles, which are shock heated to roughly $10^6$
  K, and then slowly cooled down somewhat. These constitute less than
  $15\%$ of the particles with $T> 10^{4.5}$ K.
\item{\bf 3)} Cold particles close to the centre ($r<15$ kpc). These
  are shock heated, and then they cooled down very quickly. These may
  have experience several episodes of heating and cooling.
\item{\bf 4)} Cold particles which are never shock heated. These
  particles come through the cold accretion phase.
\end{itemize}

The two cold modes (3 and 4 above) 
were discussed in detail in \cite{2005MNRAS.363....2K,2006ApJ...636L..25M}.

\begin{figure}[htb]
	\centering
	\includegraphics[angle=0,width=0.45\textwidth]{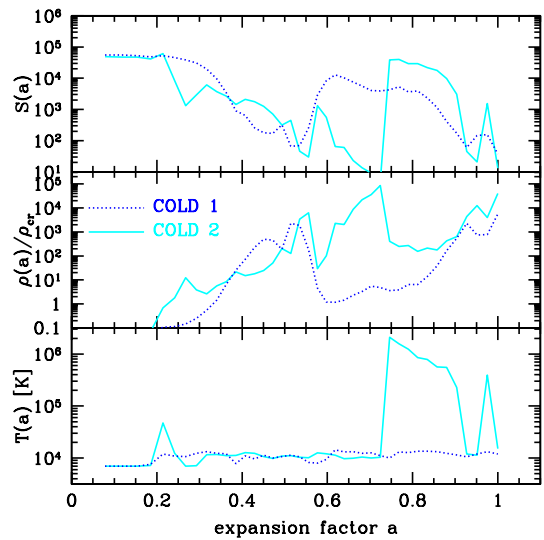}
	\caption{History of 2 cold gas particles. Shown is entropy, local
          density, and temperature as function of expansion factor.}
\label{fig:coldnew}
\end{figure}

In Figs.~(\ref{fig:hothot},\ref{fig:coldnew}) we present the entropy,
local density near the particle, and temperature as function of the
expansion factor.  In Fig.~{\ref{fig:hothot}} we present the history
of 2 particles which are hot today.  The first particle (Hot 1) 
only has one period of heating as it enters the final halo.  The
second gas particle (Hot 2) has one episode of heating followed by rapid
cooling in its first subhalo, followed by heating when it enters the
final halo.
%
In Fig.~{\ref{fig:coldnew}} we present the history of two particle
which are cold today. 
The first (Cold 1)
has been accreted
cold, and it never experienced any shock heating.
The second particle (Cold 2)
had 2 periods 
of heating followed by rapid cooling, which is clearly seen to 
happen when the particle enters the high density region.

\subsection{Evolution of $\kappa$ with redshift}

In order to test the generality of the above results, that $\kappa(r)
\approx 1$, we present results from independent simulations of 
\cite{2008aApJ...685L.105R,2008bApJ...687L..13R,2009ApJ...702.1250R}
using the parallel version of the hybrid SPH/$N$-body FTM 4.5 code
(Heller \& Shlosman 1994; Heller, Shlosman \& Athanassoula 2007). The
gravitational forces are computed using the falcON routine
(Dehnen 2002) which scales as O(N). The tolerance
parameter $\theta$ is fixed at 0.55. The gravitational softening
applied is $\epsilon = 500$~pc for the DM, stars, and gas.  The
vacuum boundary conditions are used and the simulations are
performed with physical coordinates. The cosmological constant is 
introduced by an explicit term in the acceleration equation. The
conservation of the total angular momentum and energy within the
computational sphere in the colllisionless models is within $\sim
0.01\%$ and $\sim 1\%$, respectively. The evolution of various
parameters characterizing the DM and baryons is followed in 1000
snapshots, linearly spaced in the cosmological expansion parameter
$a$.

The modeling of star formation (SF) processes and associated feedback
are described in Heller \& Shlosman (1994) and Heller et al. (2007),
which should be consulted for details. Feedback from OB stellar winds
and supernovae (SN) Type II is also included. A fraction of this
energy is thermalized and deposited in the gas in the form of
thermal energy, then converted to kinetic energy through the equations
of motion.  This method is preferable over injecting a fraction of the
stellar energy directly in the form of a kinetic energy (Heller et al. 2007).

Multiple generations of stars are allowed to form from each gas
particle. The evolution of gas metallicity is followed and the
fraction of massive stars that lead to the OB stellar winds and SN is
calculated from the Salpeter IMF.

\begin{figure}[thb]
	\centering
	\includegraphics[angle=0,width=0.45\textwidth]{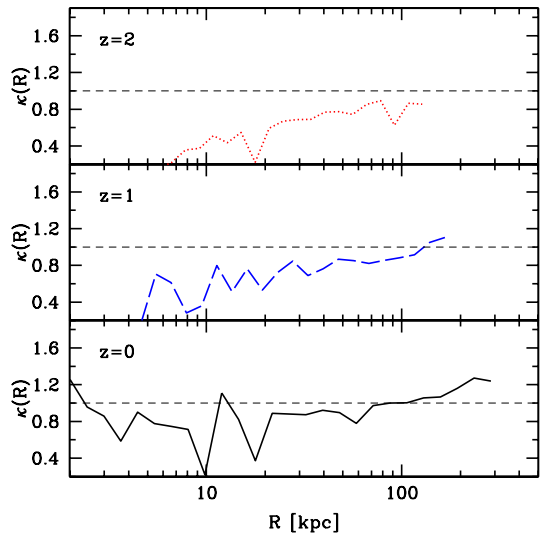}
	\caption{Evolution of $\kappa(r)$ with redshift using
          cosmological simulations of galaxy formation and evolution
          by Romano-Diaz et al. (2009). From top to bottom the frames
show $\kappa(r)$ at redshifts $z=2,1,0$.}
\label{fig:kappaIY}
\end{figure}


Fig.~\ref{fig:kappaIY} displays the redshift evolution of the radial
profiles of the temperature ratio, $\kappa$. 
All the gas particles (both hot and cold) are included
in the averages here, and therefore different aspects of 
the time evolution of $\kappa (r)$
can be followed.  At higher redshifts, $z \geq 2$, when
the disk is being assembled from the cold gas supplied by the
filaments, and which is converted into stars, the temperature ratio
drops below unity in the disk region. For $z < 1$, when the star
formation, feedback by SN and stellar winds (Romano-Diaz et al. 2008b,
2009) consume a large fraction of this gas, the temperature ratio
rises to unity, as the heated gas tends to the DM
temperature. After $z \approx 1$ we observe an influx of subhalos along
the filaments which ablate the cold gas from the disk nearly
completely. Two independent parameters verify the overall evolution of
the model from a late-type to the early-type: the fraction of the
gas-to-disk mass ratio, and the ratio of a spheroidal-to-disk mass
ratio (within the disk radius). At $z=0$ we find a temperature ratio
which is slightly below unity in the central region, and rises to 
slightly above unity towards the virial radius.

Hence, we see that the gas-to-DM temperature ratio
approaches unity for the evolved and relaxed galaxies. Of course, the
actual evolution of the ratio does depend on the particular history of
the galaxy, but the important point here is, that for equilibrated
galaxies today, the temperature ratio is close to unity.

As a sidenote we point out that the stellar dispersion does not 
agree with the DM dispersion for these galaxies. We can therefore
not use the stellar velocities in the same way to extract information
about the DM, and we cannot assume that the stellar velocity dispersion
anisotropy should equal that of the DM.

\section{AGN outflows}
\label{sec:agn}

The simple relation between the gas and DM velocity dispersions in
eq.~(\ref{eq:sigtot}) has been demonstrated using numerical
simulations for galaxy clusters in \cite{host2009}.  Those simulations
had sufficient resolution to probe a region outside of roughly
100-200~kpc. It remains unknown if this simple relation still holds
further towards the central region, where cooling may be faster, or where a
central AGN may pump energy into the intracluster gas.

To address this question, we present results of a simulation of a
three-dimensional model of AGN self-regulation in a cool-core cluster
\citep{2009MNRAS.398..548B}. 
This simulation was performed with  FLASH version 3.0 \citep{2000ApJS..131..273F}
a multidimensional adaptive mesh refinement hydrodynamics
code, which  solves the Riemann problem on a Cartesian grid using a
directionally-split  Piecewise-Parabolic Method (PPM) solver.
In this simulation the cluster properties, such as the
density and temperature profiles have been chosen to resemble the
best-studied cluster, the Perseus cluster. The gravitational potential
is taken to be static, and initially the gas is set up in hydrostatic
equilibrium. The gas physics includes radiative cooling, heating
through AGN feedback and a subgrid model for Rayleigh-Taylor-driven
turbulence, as described in \cite{2009MNRAS.398..548B}. Unlike in
previous simulations, the energy input rate was not predetermined, but
instead calculated from the instantaneous conditions near the centre
of the cluster. A fixed fraction of the $mc^2$ rest mass
energy of the accreted gas is returned to the ICM in the form of
AGN-blown cavities whose turbulent evolution couples them to the
inflowing cool gas \citep{2009MNRAS.395.2210B}.  
When turbulence is properly accounted for, the
model 
results in a
self-regulated AGN, where a phase of an episodic heating is followed by
a quiescent phase, where the cool-core reforms.

For a wide range of feedback efficiencies, the cluster regulates
itself for at least several $10^9$ years. Heating balances cooling
through a string of outbursts with typical recurrence times of around
80 Myrs, a timescale that depends on global cluster properties.

\begin{figure}[thb]
           \centering
           \includegraphics[angle=0,width=0.45\textwidth]{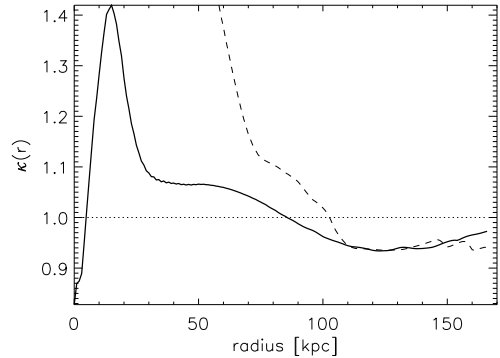}
           \caption{The temperature ratio as function of radius for
             the simulated galaxy cluster with a recurrent AGN heating
             source. The full line shows the temperature ratio of the
             fiducial run of a self-regulating AGN in a cool core
             cluster after 1 Gyr, labeled D5-10 in \cite{2009MNRAS.398..548B}.  This
             figure demonstrates that outside the central region
             ($\approx 30$ kpc, where the recent AGN burst has heated the
             gas temperature much above the DM temperature) the
             ratio $\kappa (r) = \sigma^2_{\rm baryons} / \langle
             \sigma^2_{\rm DM} \rangle$ is unity, within $10\%$. The
             dashed line shows the results from an extreme case of
             AGN-feedback, where strong outflows occur
             every 50 Myrs \citep{2008ApJ...686..927S}.  }
    \label{fig:kappaMB}
\end{figure}

In Fig.~(\ref{fig:kappaMB}) we see that for the self-regulating AGN
feedback model after 1 Gyr (solid line), the spherically averaged gas
dispersion does indeed agree with the DM dispersion within $10\%$, at
radii larger than 30 kpc.  This case represents the best available
simulation of a self-regulating cool core cluster which reproduces a
number of key features seen in X-ray observations
\citep{2009MNRAS.398..548B}.

In the central region, at radii smaller than 30~kpc, we see a
sharp increase in the gas temperature resulting from a recent AGN
burst, and at the very centre, within less than 10~kpc, we observe the
effect of very rapid gas cooling, and the re-emergence of a cool core.

To test an extreme case of AGN heating of a cluster core, we also
present the results from a model where AGN-driven bubbles occur at
regular intervals of 50 Myrs with bubble energies of $5 \times
10^{59}$ ergs \citep{2008ApJ...686..927S}. Even for this non-self
regulated, and somewhat less physically motivated case, the dashed
line in Fig.~(\ref{fig:kappaMB}) shows that, as long as we consider
radii outside of 70--100~kpc region, the velocity dispersion ratio
stays within $10\%$ of unity.

\section{Implications for the velocity anisotropy}
\label{sec:discussion}

For the self-regulated AGN, the simple temperature relation $\kappa
(r)\approx 1$ holds outside the inner region of $r\approx 30$~kpc, as we
have shown in section~4.  For the observed X-ray clusters studied in
\cite{host2009}, this radius corresponds roughly to the centre of the
innermost bin. This shows that one can trust the inference of the DM
velocity anisotropy in galaxy clusters down to radii as small as $\approx
50$~kpc.

In the case of galaxies, we have demonstrated that the simple
temperature relation holds for relaxed and equilibrated structures,
when applied to radii beyond the regions dominated by the stellar
component. This opens for the possibility of indirectly determining the
DM velocity anisotropy in galaxies in the following way. By measuring
the temperature and the density of the hot X-ray emitting gas, one may
use the equation of hydrostatic equilibrium, eq.~(\ref{eq:hydro}), to
measure the total gravitating mass.  Now, considering the Jeans
equation, eq.~(\ref{eq:jeans}), we see that it contains the total mass
on the left hand side, and 3 unknown quantities on the right hand side
($\rho_{DM}, \sigma_r^2, \beta$). First, combining the total mass
measurement from X-ray with spectral determination of the stellar
mass, allows one to determine the DM density profile
\begin{equation}
\rho _{DM} \left( r \right) = f_1 \left(T_g (r), \rho_g (r), \rho_{\rm star} (r) \right) \, .
\end{equation}
We are thus left
with 2 unknowns on the right hand side
of the Jeans
equation, eq.~(\ref{eq:jeans}), namely the radial dispersion
and the velocity anisotropy, $\beta$. Using the temperature relation,
eq.~(\ref{eq:sigtot}), together with the X-ray determined gas
temperature, we get rid of one of these, and hence get an indirect
measurement of the dark matter velocity anistropy
\begin{equation}
\beta (r) = f_2 \left( \kappa (r) , T_g (r), \rho_g (r) , \rho_{\rm star} (r)  \right) \, .
\end{equation}
Such observation
has only very recently become possible for X-ray bright galaxy cluster
\citep{hansenpiff, 2007MNRAS.380.1521M, host2009}, and it is clear that such indirect
observations in galaxies most likely will require improved X-ray
satelites. The simplest possibility would be if $\kappa (r) =1$ everywhere, 
however, the method to infer the DM velocity anisotropy 
as described above, works as long as the radial form
of $\kappa (r)$ is known \citep{host2009}. This is exactly one of the points
of this paper, that the form of $\kappa (r)$ appears to be universal for
galaxies (see figure \ref{fig:kappaAM}).

A direct measurement of this velocity anisotropy is expected to be
virtually impossible in terrestrial experiments
\citep{hosthansen2007}.  Measuring $\beta$ of the DM in galaxies, as described
above, can provide an independent confirmation to virtually every
numerical modeling of DM halo formation in the cosmological context,
and to the suggestion that a non-zero $\beta$ is a fundamental
property of galactic DM halos,
even in equilibrium \citep{2009ApJ...694.1250H,2010ApJ...718L..68H}.  
Furthermore, a non-zero $\beta$ has an effect on
the underground direct detection experiments
\citep{2000PhRvD..62b3519V,2000MNRAS.318.1131E,2008PhRvD..77b3509V}.
A detection through the method laid out above would therefore decrease
the systematic error-bars in terrestrial DM detection experiments.

\section{Conclusions}
\label{sec:conclusion}
In summary, we have found a very simple relation between the
temperature of the hot gas and the averaged dispersion of the dark
matter, namely that their ratio is close to unity, $\kappa \approx 1$.
We have demonstrated that this near equality of the gas and DM
temperatures holds in the case of equilibrated and relaxed
galaxies. 

This relation is not only conceptually important, but it
will also allow an indirect determination of the dark matter velocity
anisotropy in galaxies. Such an observation will provide an important
and independent confirmation of all cosmological DM halo formation
simulations.

We have also studied this relation in galaxy clusters containing
an active AGN, and our results confirm that one can indeed use
galaxy clusters to infer the dark matter velocity anisotropy.

\newpage 
\noindent
It is a pleasure to thank Isaac Shlosman for 
insightful discussions.
SHH thanks Ole Host, Marco
Roncadelli and Darach Watson for discussions.  
Numerical simulations were performed on
the PIA and on the PanStarrs2 clusters of the Max-Planck-Institut
f\"ur Astronomie at the Rechenzentrum in Garching and on the zBox2
supercomputer at the University of Z\"urich.  Special thanks to
B. Moore, D. Potter and J. Stadel for bringing zBox2 to life.
E.R.D. simulations
have been run on a dedicated cluster.
The AGN simulations were conducted on the "Saguaro" cluster operated by 
the Fulton School of Engineering at Arizona State University.
Y.H. has been partially
supported by the ISF (13/08).  
The Dark Cosmology Centre is funded by
the Danish National Research Foundation.


\end{document}